\documentclass[a4paper,11pt]{article}

\usepackage{jinstpub} 

\usepackage{lineno}

\usepackage{url}
\usepackage{multirow}

\usepackage{subcaption}
\usepackage{gensymb}
\usepackage{comment}
\usepackage{bm}

\begin{document}

\title{Validation of the VUV-reflective coating for next-generation liquid xenon detectors}

\author[1]{D.~Bajpai}
\author[1]{A.~Best}
\author[1,2,a]{I.~Ostrovskiy}\note[a]{Corresponding author: iostrovskiy@ua.edu}
\author[3]{D.~Poitras}
\author[1]{W.~Wang}

\affiliation[1]{Department of Physics and Astronomy, University of Alabama, Tuscaloosa, Alabama 35487, USA}
\affiliation[2]{Currently at Institute of High Energy Physics, Beijing, China}
\affiliation[3]{Quantum and Nanotechnologies Research Centre, National Research Council of Canada, Ottawa, Ontario K1A 0R6, Canada}
\abstract{Coating detector materials with films highly reflective in the vacuum ultraviolet region improves sensitivity of the next-generation rare-event detectors that use liquid xenon. In this work, we investigate the MgF$_2$-Al-MgF$_2$ coating designed to achieve high reflectance at 175 nm, the mean wavelength of liquid xenon (LXe) scintillation. The coating was applied to an unpolished, passivated copper substrate mimicking a realistic detector component of the proposed nEXO experiment, as well as to two unpassivated substrates with ``high'' and ``average'' levels of polishing. After confirming the composition and morphology of the thin-film coating using TEM and EDS, the samples underwent reflectance measurements in LXe and gaseous nitrogen (GN2). Measurements in LXe exposed the coated samples to -100 $\degree$C for several hours. No peeling of the coatings was observed after several thermal cycles. Polishing is found to strongly correlate with the measured specular reflectance ($R_{\mathrm{spec}}$). In particular, 5.8(5)\% specular spike reflectance in LXe was measured for the realistic sample at 20$\degree$ of incidence, while the values for similar angles of incidence on the high and average polish samples are 62.3(1.3)\% and 27.4(7)\%, respectively. At large angles (66\degree--75$\degree$), the $R_{\mathrm{spec}}$ in LXe for the three samples increases to 23(5)\%, 80(8)\%, and 84(18)\%, respectively. The $R_{\mathrm{spec}}$ at around 45$\degree$ was measured in both GN2 and LXe for average polish sample and shows a reasonable agreement. 
Importantly, the total reflectance of the samples is comparable and estimated to be 92(8)\%, 85(8)\%, and 83(8)\% in GN2 for the realistic, average, and high polish samples, respectively. This is considered satisfactory for the next-generation LXe experiments that could benefit from using reflective films, such as nEXO and DARWIN, thus validating the design of the coating.}

\keywords{Dark Matter detectors (WIMPs, axions, etc.); Double-beta decay detectors; Noble liquid detectors (scintillation, ionization, double-phase); Time projection chambers;}

\maketitle

\flushbottom

\section{Introduction}
\label{sec:intro}
Next-generation rare-event searches with liquid xenon (LXe) detectors plan to utilize tons of LXe and achieve unprecedented sensitivity to WIMP dark matter~\cite{darwin2016}, neutrinoless double-beta decay~\cite{nexo,darwin0nu} (0$\nu\beta\beta$), and other physics channels~\cite{darwin_solarNu,g3}. In particular, nEXO~\cite{nEXO_pCDR} is a proposed 5-ton detector aimed to achieve $\sim$10$^{28}$ yr half-life sensitivity to 0$\nu\beta\beta$. nEXO is loosely based on a much smaller EXO-200 experiment~\cite{exo_jinst}. Like EXO-200, it is planning to construct a single-phase, cylindrical time projection chamber (TPC) filled with xenon enriched in $^{136}$Xe isotope. Otherwise, nEXO's design is different and based on novel approaches to charge and light readout. Instead of EXO-200's charge-collecting anode wires, nEXO envisions the anode as an array of small metal pads deposited on fused-silica substrates~\cite{pads}. 
For the light readout, EXO-200 used photosensors located behind the charge-collecting wires at the end caps of the TPC. The barrel of the TPC was covered with PTFE panels with high VUV diffuse reflectance that were placed in front of the copper field-shaping rings (FSRs). Instead, nEXO will have to remove the PTFE panels in order to place photosensors behind the FSRs, since the charge readout pads, and the copper cathode, are planned to be nontransparent. This approach would substantially reduce the light collection efficiency (LCE) due to the obstruction by the FSRs and their poor reflectivity in the VUV. Consequently, nEXO is planning to increase the reflectivity of the FSRs and cathode by coating them with VUV-reflective films. Based on simulations, the FSRs and cathode need to be made 80\% reflective to achieve the experiment's goals~\cite{nexo}. 

Another next-generation LXe rare-event search, DARWIN~\cite{darwin2016}, aims at a tenfold increase in sensitivity to WIMPs, compared to the current experiments~\cite{XENONnT,LZ}. It plans to instrument $\sim$40 tons of natural xenon in a dual-phase TPC -- a technology that gradually matured and has demonstrated ton-level viability~\cite{xenon1t_instrumental, LZ_instrumental,panda_x}, a keV-level energy threshold~\cite{xenon1t_s2only,lux_thr}, and sub-percent energy resolution at MeV energies~\cite{xenon1t_eres}. A competitive measurement of solar neutrinos~\cite{darwin_solarNu} and 0$\nu\beta\beta$~\cite{darwin0nu_erratum} will also be possible, the latter thanks to having a similar mass of $^{136}$Xe to the dedicated experiment~\cite{nexo} without the need for enrichment. In the default approach to light collection~\cite{darwin2016}, DARWIN will use the highly reflective PTFE sidewalls and electrodes with $\gtrsim$90\% transparency. The scintillation light will be collected with photosensors placed at the end caps of the TPC. While DARWIN is already expected to have LCE that is large enough to achieve its goals, the reflective coatings may provide some extra advantage. As was shown~\cite{darwin_chroma}, coating the DARWIN electrodes with reflective films is expected to increase the LCE by up to 24\% rel. 

Motivated by the above considerations, an effort to develop a highly VUV-reflective mirror coating has been performed recently~\cite{china_mirror}. The design of the coating was based on an aluminum layer deposited by thermal evaporation in vacuum. Aluminum's reflectivity in the VUV is known to be one of the highest~\cite{r_vuv_1939}. To protect from oxidation, the aluminum was coated by a MgF$_2$ layer. Additionally, it was found that a layer of SiO$_2$ was necessary between the metal substrate and aluminum to preserve the high reflectance that was negatively affected by the alloying of the two metals. The design achieved the specular reflectance above 80\%, as measured in vacuum at 175 nm, the mean wavelength of scintillation in liquid xenon~\cite{lxe_wave}. However, two issues remain to be addressed to confirm the applicability of the coating to the next-generation experiments. Firstly, the surface treatment of realistic detector components is substantially different from the idealized substrates used in the above study. In particular, the copper substrates were highly polished to get the surface roughness of just a couple of nm, while the nEXO's FSRs and cathode surfaces are not planned to be polished, partly due to the risk of compromising the radiopurity. Additionally, a chemical passivation treatment is foreseen for the detector's copper surfaces. It is not clear how the surface roughness and chemical treatment may affect the adhesion, functionality of the alloy barrier film and the coating's reflectance. Secondly, strict radiopurity goals require assessment of all additional components and selection of most radiopure stocks of materials used in the film deposition, leading to a preference for the minimal number of different coating components. This work describes the characterization of reflective coating samples prepared to address both of the issues. The samples are made a) with a MgF$_2$ alloy barrier layer instead of SiO$_2$ and b) on substrates with different surface treatments, including the one foreseen for nEXO. The following sections will describe the samples, reflectance measurements, and results of the study.

\section{Description of the samples}
\label{sec:samples}
\begin{figure*}[htpb]
    \centering
    \includegraphics[width=0.32\textwidth]{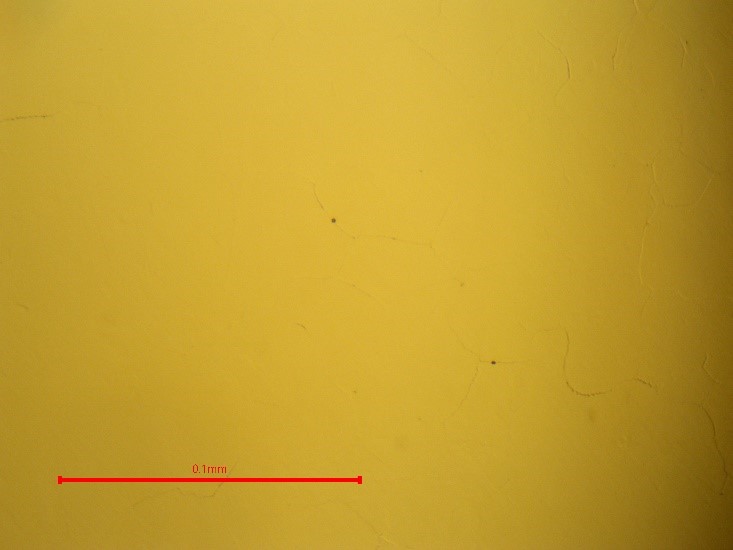}
    \includegraphics[width=0.32\textwidth]{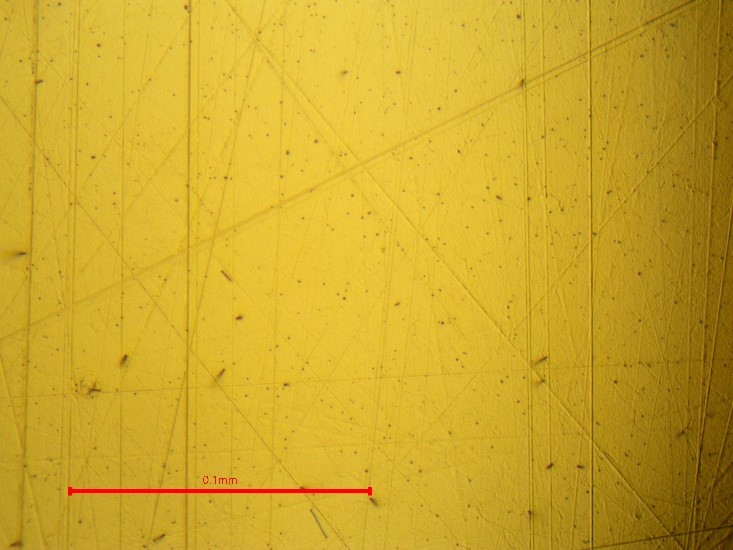}
    \includegraphics[width=0.32\textwidth]{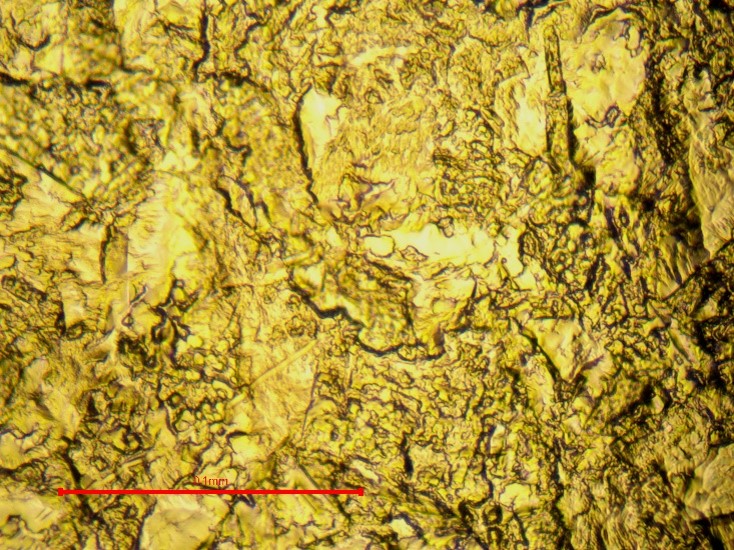}
    \caption{Optical microscope photographs of the high polish (left), average polish (center), and realistic samples studied in this work. The extent of the red lines corresponds to a 100 micron length standard. The long, micrometer-wide scratches covering the average polish substrate are unexpected.}
    \label{fig:samples}
\end{figure*}

The samples are prepared by consecutively depositing films of MgF$_2$ ($\sim$20\,nm), aluminum ($\sim$100\,nm), and MgF$_2$ ($\sim$100\,nm) on copper substrates, starting from the thin MgF$_2$ layer. The details of the deposition are the same as in Ref.~\cite{china_mirror}. The substrates are disks of 1 inch diameter and 0.1 inch thickness that have different types of surface treatment. The ``high'' polish substrate is prepared the same way as described in Ref.~\cite{china_mirror} by the Institute of Optics and Electronics, CAS, China, to the $\sim$2\,nm surface roughness. The ``average'' polish substrate is mechanically polished by Valley Design, USA~\cite{valley}, to the measured $\sim$10\,nm surface roughness. The ``realistic'' substrate is manufactured using a 0.25-inch ball mill with a 0.005-inch step at the Pacific Northwest National Laboratory, USA, and is not mechanically polished. The realistic substrate has been chemically passivated according to the protocol described in Ref.~\cite{passivation}. Figure~\ref{fig:samples} shows optical microscope photographs of the surfaces of the three samples. 
The appearance of the samples is generally in accordance with the expectation based on their corresponding surface treatment, with the exception of the average polish substrate that exhibits long, micrometer-wide scratches across its whole surface. 

\begin{figure*}[htpb]
    \centering
    \includegraphics[width=0.225\textwidth]{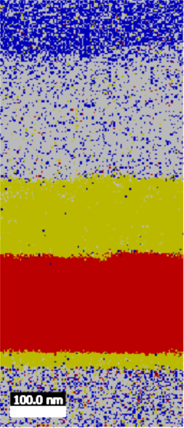}
    \caption{EDS image of a specimen lifted out from a high polish sample confirming the expected composition and thickness of the layers. The region with identified K-shell emissions of Mg and F elements is colored yellow. Red corresponds to the Al K-shell emissions. Blue represents Cu (substrate) and Pt (FIB LO-related contamination) K-shell lines. A 100-nm length standard is shown in white.}
    \label{fig:eds}
\end{figure*}

\begin{figure*}[htpb]
    \centering
    \includegraphics[width=0.95\textwidth]{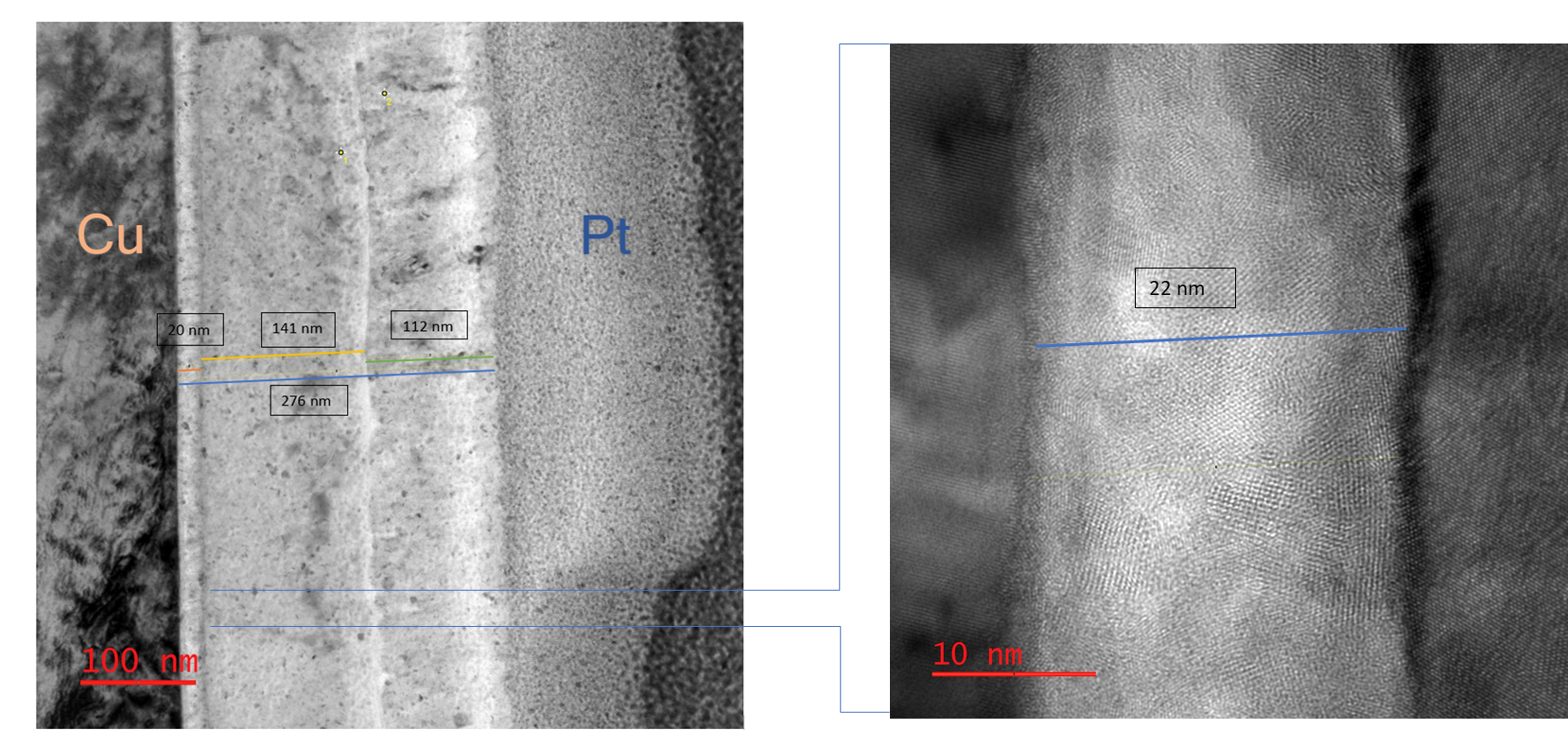}
    \caption{TEM images of a specimen lifted out from a high polish sample. (Left) The three deposited layers are seen between the copper substrate and platinum LO layer. The thickness of the stack and individual deposited layers is also shown. (Right) A zoomed in image of the alloy barrier layer. The copper substrate and aluminum layer are clearly separated. Length standards are shown in red.}
    \label{fig:tem}
\end{figure*}

To confirm the composition of the deposited layers, one of the spare high polish samples was studied at the University of Alabama's Analytical Research Center~\cite{aarc} using an FEI Tecnai F-20 Transmission Electron Microscope (TEM) equipped with an EDAX energy dispersion spectrometer (EDS). The cross-sectional specimens were lifted out from the sample using the focused ion beam (FIB LO) technique. Figure~\ref{fig:eds} shows an EDS image of the sample's specimen, confirming the expected composition of the deposited layers. 
Figure~\ref{fig:tem} shows a TEM image of the same specimen. The thicknesses of the three layers deposited on the copper substrate (Cu) and the overall thickness of the stack are shown on the left image. 
The right image zooms in on the thin alloy barrier layer demonstrating a successful separation between the copper substrate and aluminum layer. The aluminum and top MgF$_2$ layers are slightly thicker than expected, which does not negatively affect the performance. 

\section{Reflectance measurements}
\label{sec:setups}

\subsection{LIXO2: reflectance in LXe}
\begin{figure*}[htpb]
    \centering
    \includegraphics[width=0.65\textwidth]{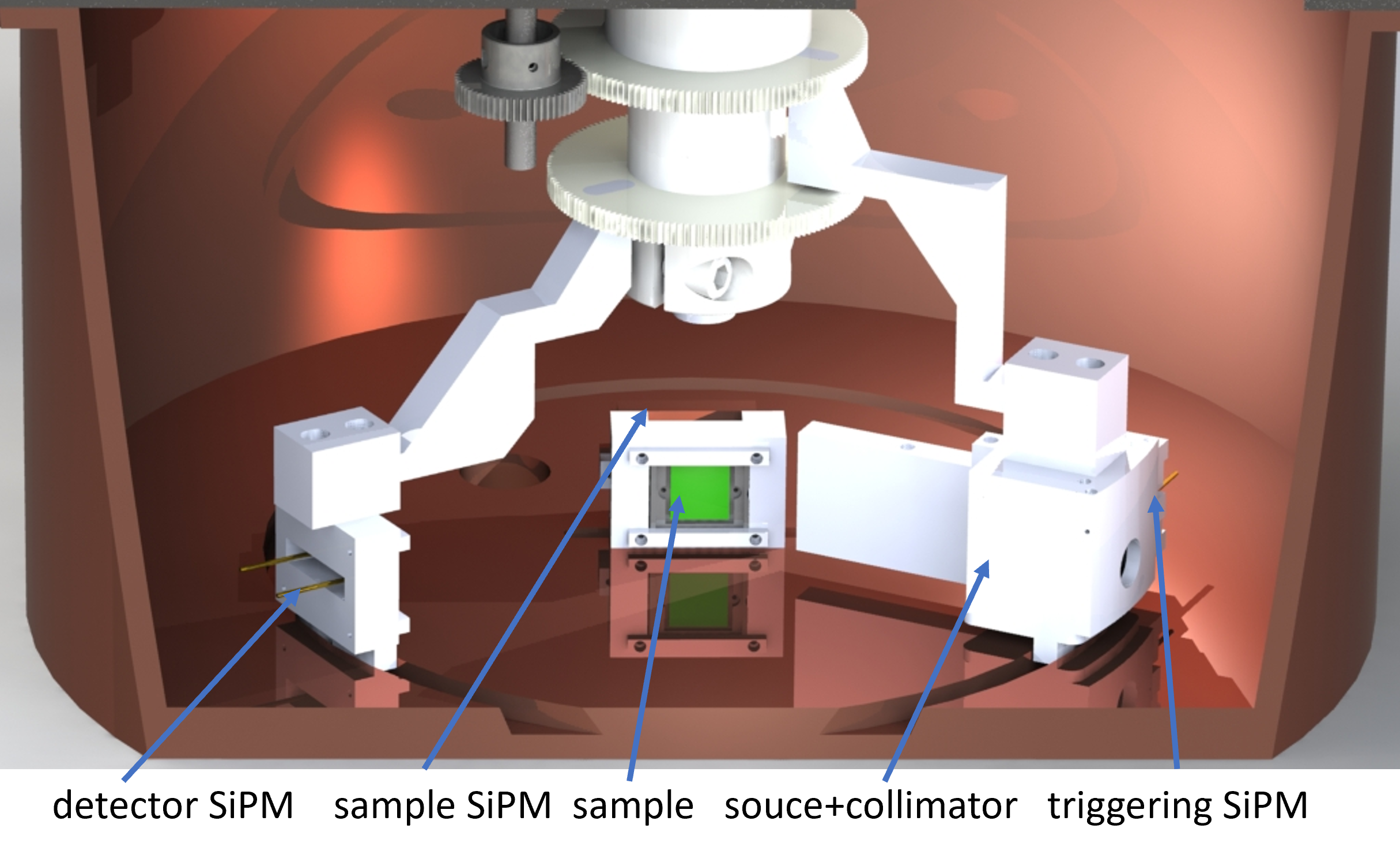}
    \caption{CAD model of the LIXO2 payload inside the LXe chamber. A reflecting (transmitting) sample is installed in a PEEK holder at the center. A SiPM photodetector rotates around the sample to measure the reflected (transmitted) scintillation light excited by a radioactive source inside a PEEK collimator. A triggering SiPM is attached to the side of the collimator for triggering and light yield stability purposes. During reflectance measurements, another SiPM is installed immediately behind the sample to monitor the stability of the light flux at the center of the chamber.} 
    \label{fig:payload}
\end{figure*}

The reflectance of the samples in LXe is measured using a setup, LIXO2, constructed at the University of Alabama. The setup shares with its earlier version, LIXO, the xenon handling and data acquisition systems, which are described elsewhere~\cite{lixo}. The LIXO2's cryostat is a larger version of the LIXO's one, with additional feedthroughs to control stepper motors. The LIXO2's LXe chamber features two ultrahigh vacuum rotary feedthroughs that transmit torque from the two external stepper motors to the payload inside the chamber using magnetically-coupled stainless-steel shafts~\cite{md10}. The payload consists of a system of stainless-steel gears and lever arms that enable the rotation of the collimated scintillation light source and reflected light detector around the sample placed in a PEEK holder at the center. The detector is a 6x6 mm$^2$ Hamamatsu VUV4 S13370-6050CN SiPM~\cite{vuv4}. The LXe scintillation light is excited with an $^{241}$Am radioactive source that is enclosed in a chamber of a PEEK collimator. The collimator has an opening of 0.99 mm diameter. The source-sample and sample-detector distances are 38.2 and 41.2 mm, respectively. Another S13370-6050CN SiPM attached to the side of the collimator such that its field of view is limited to the inside of the collimator's chamber. The collimator SiPM provides a way to trigger on the scintillation light flash caused by a $^{241}$Am decay and to monitor the stability of the light yield. A third S13370-6050CN SiPM is placed immediately behind the sample to monitor the stability of light flux at the same distance from the source as the sample. During each measurement, the collimated source is placed behind a sample (at 180$\degree$ of incidence) directly aimed at the sample SiPM. The light seen by the sample SiPM may in principle vary independently from the light seen by the collimator SiPM due to changes in the LXe absorption length. During the whole measurement campaign, the collimator's and sample's SiPMs measured the light levels to be stable to within 3\% rel. and 4\% rel., respectively. The measured variation was used as a correction to the light seen by the detector SiPM during different measurements. Figure~\ref{fig:payload} shows a CAD model of the payload.

\begin{figure*}[htpb]
    \centering
    \includegraphics[width=0.49\textwidth]{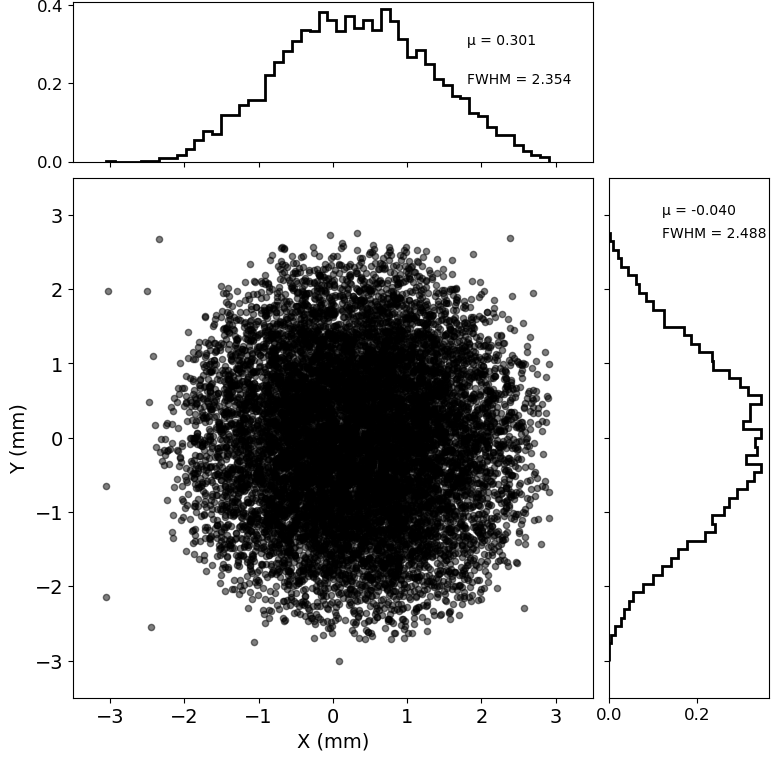}
    \includegraphics[width=0.49\textwidth]{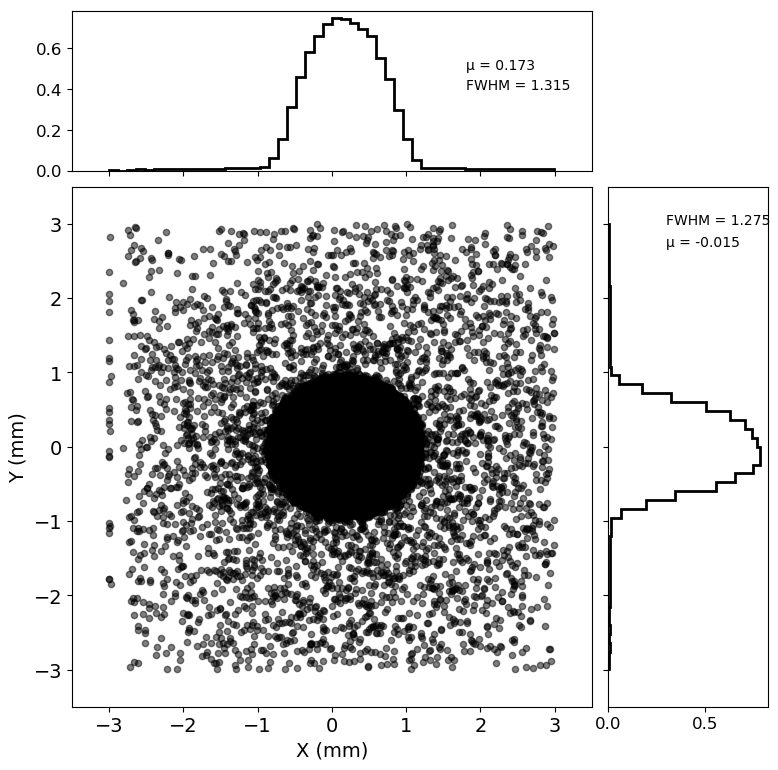}
    \caption{Distribution of photon hits on the detector's (left) and a 6x6 mm$^2$ sample's (right) positions as simulated by Chroma. The detector and collimated source are located at $\pm$20 degrees of incidence on the sample. The MC includes a detailed description of the setup's geometry and optical properties. The beam size becomes wider due to geometrical divergence and Rayleigh scattering. The latter effect seen on the detector hits distribution as occasional photon hits away from the beam spot. The larger contribution of hits away from the beam spot on the sample is mostly due to photons reflected back from the surface of the detector SiPM. The projections of the hit distributions on the x and y axes, together with their means and FWHMs, are also shown.} 
    \label{fig:beam}
\end{figure*}

\begin{figure*}[htpb]
    \centering
    \includegraphics[width=0.85\textwidth]{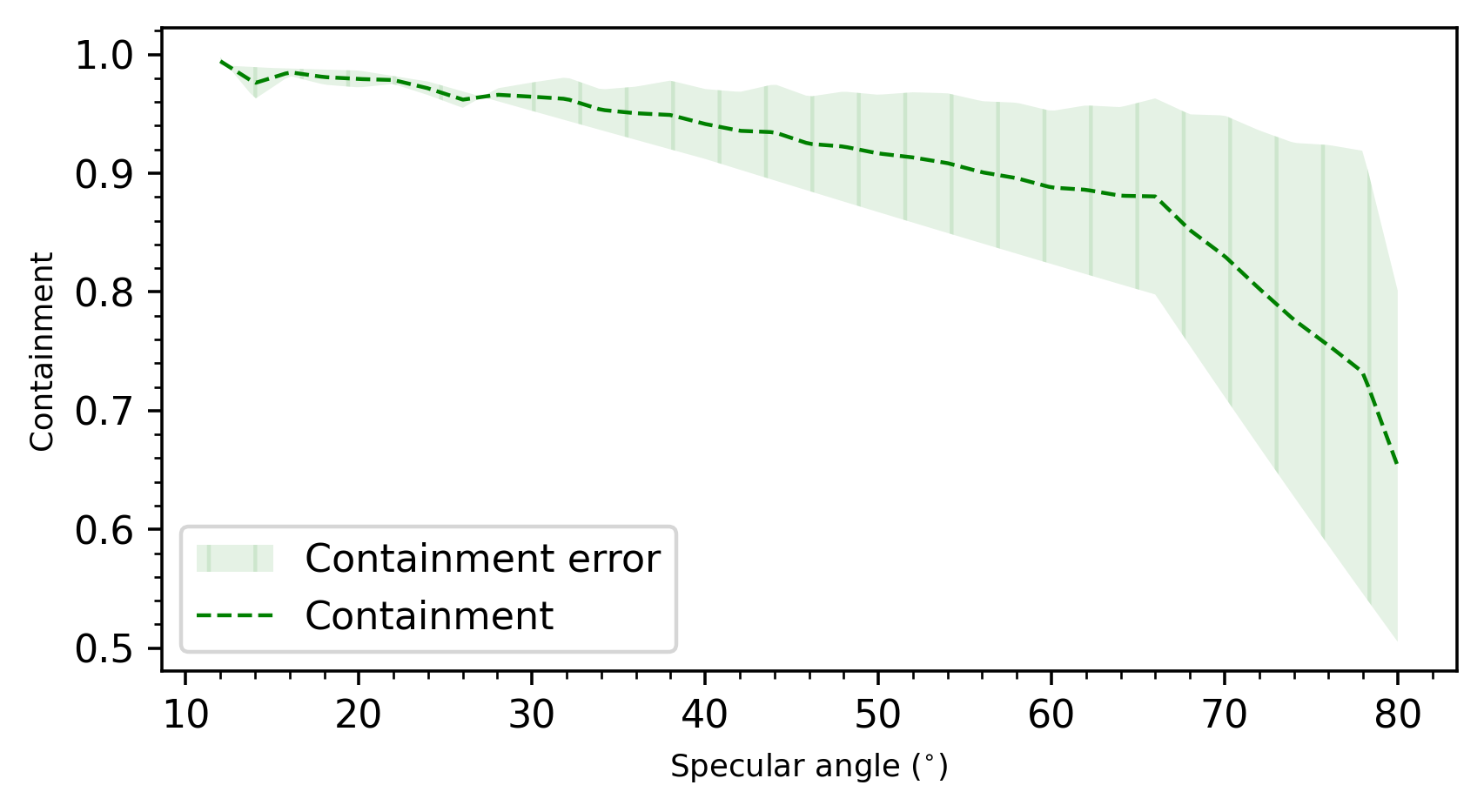}
    \caption{Containment of the light beam specularly reflected from a sample by the detector SiPM. The containment is nearly 100\% at small angles of incidence but decreases at larger angles due to shadowing by the sample holder, stretching out of the beam beyond the extent of the sample, as well as due to absorption and scattering in LXe. The shaded band represents simulation uncertainty.} 
    \label{fig:containment}
\end{figure*}

An optical simulation of the setup is implemented using the same software and approach as in Ref.~\cite{darwin_chroma}. The Monte Carlo (MC) simulation includes a detailed description of the setup's geometry and optical properties, so is expected to accurately predict the width of the light beam emitted from the 0.99-mm opening of the collimator as it gets wider due to the geometric divergence and Rayleigh scattering in LXe~\cite{ray_2}. It currently implements a sample's surface as an ideal plane, which ignores the effect of microfacets producing a specular lobe. Figure~\ref{fig:beam} shows the simulated beam profiles as distributions of photon hits on a sample and the detector SiPM for the case when the collimated source is positioned at 20$\degree$ of incidence on the sample. 
The rare photon hits seen outside of the main beam on the detector are caused by Rayleigh scatters. The out-of-beam hits are more prominent on the sample and are mostly due to the photons reflected back from the detector SiPM, which is known to have $\sim$25\% reflectivity~\cite{lixo}. This effect does not affect the samples' reflectance measurements. For the realistic assumptions on the LXe absorption length, practically 100\% of photons exiting the collimator at normal incidence on the sample hit the sample. An important parameter of the setup is the fraction of these photons that would hit the active area of the detector SiPM if the sample was a perfect specular reflector. This fraction, which we call the containment fraction, is nearly 100\% for small angles of incidence but decreases to 65\% when the collimator and detector are located at $\pm$80$\degree$ from the normal. The loss in containment is due to several factors, such as absorption and scattering in LXe, the beam becoming elliptical at large angles and extending beyond the sample, and shadowing by the sample holder. Figure~\ref{fig:containment} shows the containment fraction as a function of the angle. The band reflects the estimated uncertainty in the simulation, including a conservative contribution due to the lack of quantitative information about xenon purity, which leads to imperfectly known optical parameters of LXe~\cite{lixo}. This dependence is used to correct the measured specular reflectance at large angles. 

\subsubsection{Specular reflectance in LXe}
\begin{figure*}[htpb]
    \centering
    \includegraphics[width=0.85\textwidth]{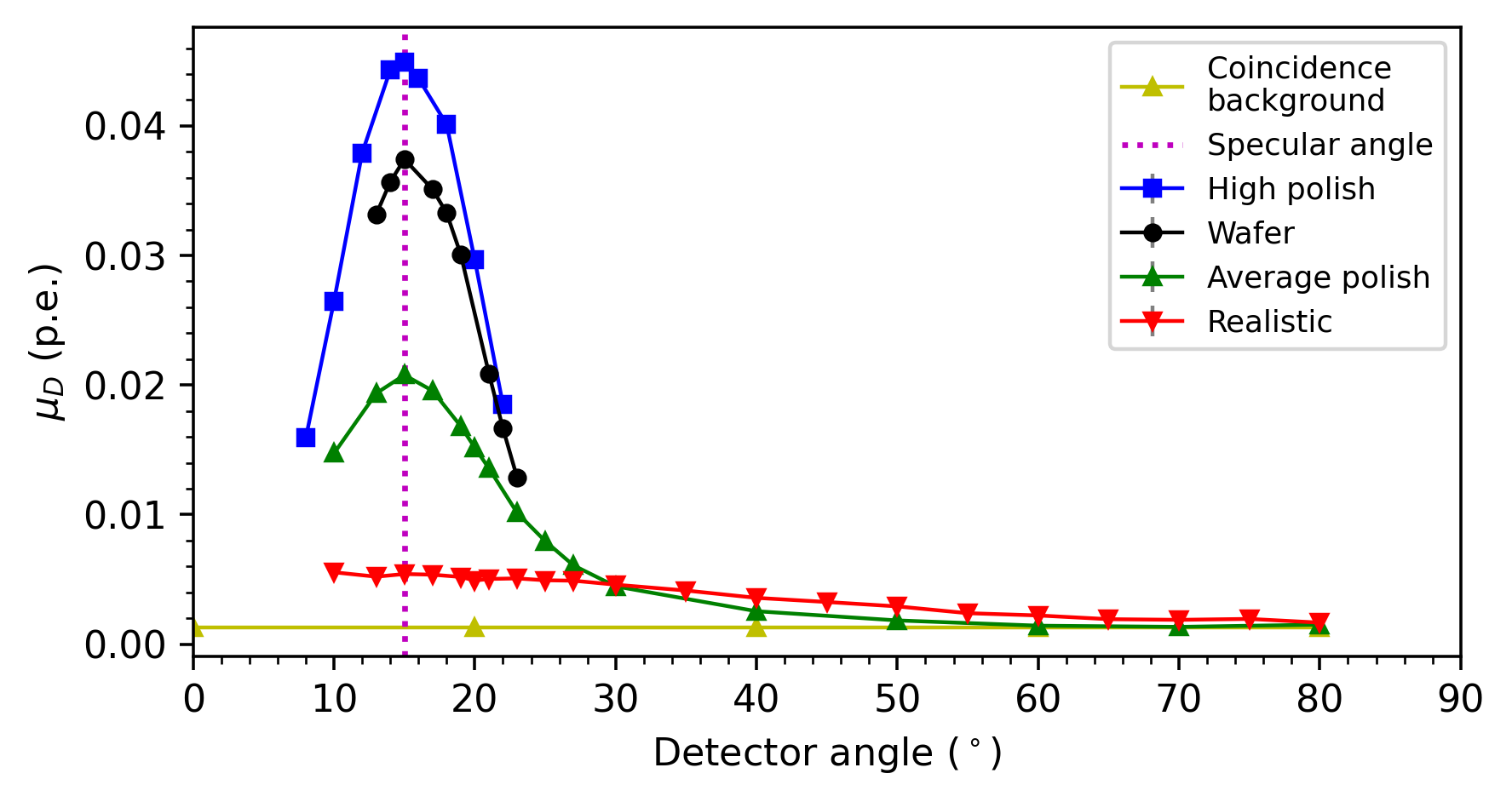}
    \caption{Mean number of p.e. seen by the detector SiPM as a function of the detector's angular position for the case when the collimated source is placed at a low (17-24) degree of incidence on the high polish (blue), average polish (green), and realistic (red) samples. The realistic and average polish samples are dominated by the specular lobe. The distribution for the wafer sample and coincidence background are also shown. The wafer sample is expected to have 51\% specular reflectance at small incidence angles and is used as a reference. The distributions are shifted to a common angular position to ease the comparison.} 
    \label{fig:small_angle}
\end{figure*}

The specular reflectance measurement method relies on the same approach of detecting coincidences between the collimator and detector SiPMs described previously~\cite{lixo}. However, extracting the absolute reflectance value is not as straightforward in this work, because the current samples have large non-specular reflectance components. Figure~\ref{fig:small_angle} shows the distributions of light reflected from the samples as a function of the detector's angular position for the case when the collimated source is placed at a low angle of incidence on the sample. The actual angles vary slightly (Table~\ref{tab:rspec}) for different samples, so the data on the figure are shifted to a common representative angular position for ease of comparison. The widths of the distributions are much larger than the angular extent of the detector SiPM (8.3\degree) for the realistic and average polishing samples. The width for the high polish sample ($\sigma\sim$4.5\degree) is comparable to the detector's width but is still larger than the beam size predicted by the MC simulation (2.4\degree\, FWHM), indicating a substantial lobe reflectance component. The width does not depend appreciably on the angular position.

The distributions for the realistic and average polish samples are dominated by the specular lobe but also show a sign of the Lambertian contribution, seen as a constant component above a background at angles far away from the specular one. However, due to the very low corresponding light levels the statistical significance accumulated so far is not enough to conclusively measure the Lambertian component in LIXO2 and will be a subject of future investigations.

The figure also shows the distribution for a piece of a silicon wafer with a 0.0015 mm SiO$_2$ layer. The wafer's VUV specular reflectance was measured in vacuum to extract the refractive index of the SiO$_2$ layer~\cite{ihep_wafer}. Together with the known layer's thickness, this allows one to use the Fresnel equations to predict the wafer's reflectance in LXe. At small angles of incidence, the wafer's reflectance is equal to $\sim$51\% and is almost insensitive to uncertainties on the indices of refraction of the LXe and MgF$_2$. At large angles the reflectance is expected to increase, reaching 100\% above the critical angle, but the rate of increase is sensitive to the indices' values, so only the low-angle wafer measurement is used as a reference reflectance in LXe in this work.

Also shown on the figure is the coincidence background that we measure \textit{in situ} by placing the collimated source behind a sample while the detector performs the angular scan in front of the sample. This background is stable during the measurement campaign to within 3\% rel. and is independent of the type of the sample. The background was also seen during previous measurements~\cite{lixo} and appears to be caused by some isotropic source of parasitic light in LXe, such as luminescence of passive components, natural radioactivity, or cosmic rays. 

The difference between the source position and best-fit mean of the distribution is indicative of the relative angular position error. It is typically better than 2$\degree$. Notably, the difference increases to 7$\degree$ at high angles of incidence for the samples with substantial non-specular reflectance. This appears to be the same effect that has been observed in Ref.~\cite{berkeley_r}. The uncertainty on the absolute angular position of the source (as opposite to the angular position of the detector relative to that of the source, discussed above) is estimated \textit{ex situ} by examining the final location of the source after each liquefaction-recovery cycle and is typically $\lessapprox$2$\degree$. 

\begin{figure*}[htpb]
    \centering
    \includegraphics[width=0.99\textwidth]{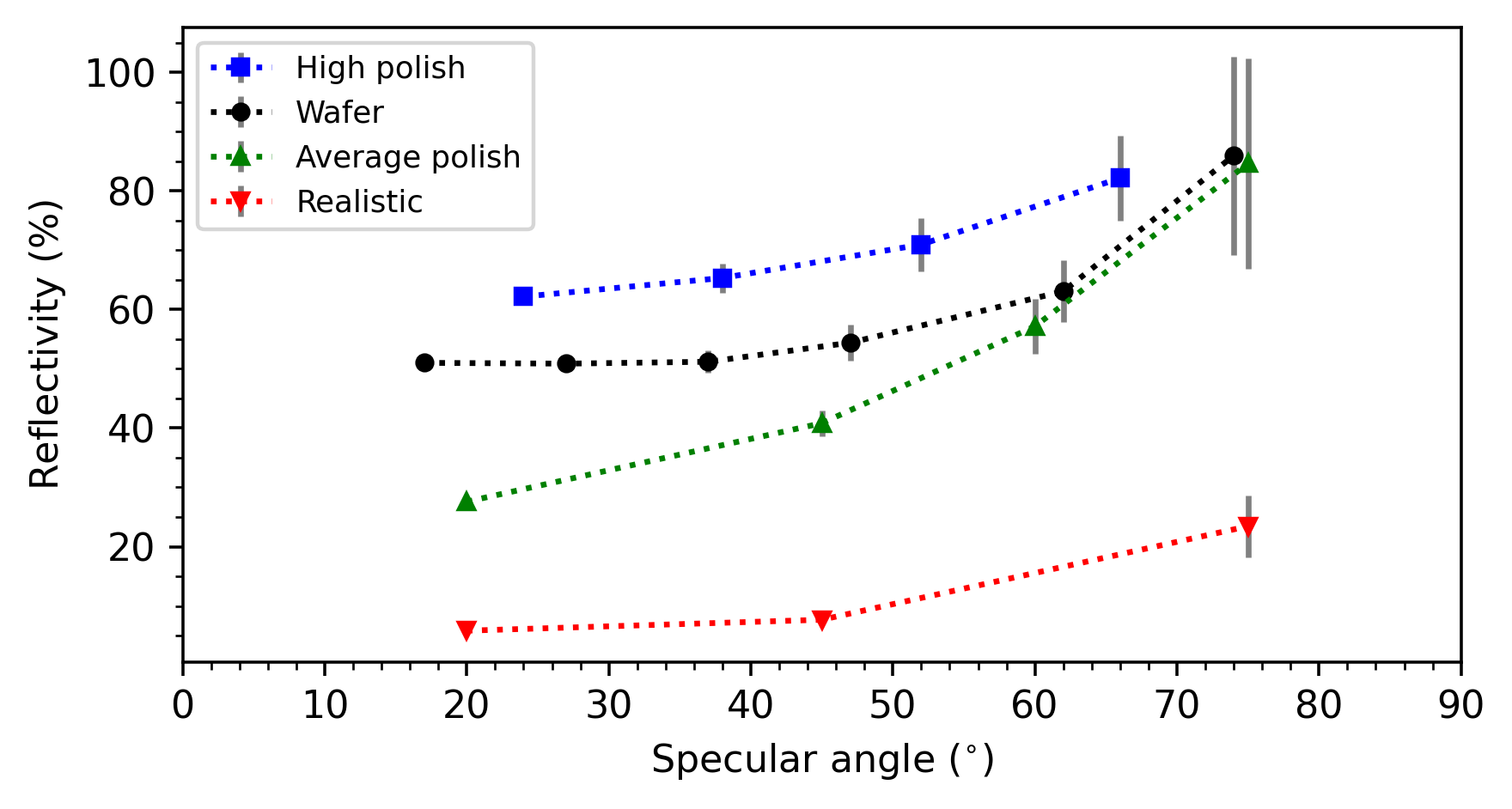}
    \caption{Specular reflectance in LXe of the samples as a function of the angular position of the collimated source. The reflectance increases with the angle of incidence and highly correlates with the quality of polishing. The data points are connected by dashed lines to guide the eye.} 
    \label{fig:rspec_vs_aoi}
\end{figure*}

To estimate specular reflectance, $R_{\mathrm{spec}}$, we use the peak value of the $\mu_{\mathrm{D}}$ (Figure~\ref{fig:small_angle}) for each angular position of the collimated source, apply corrections for the light stability and containment fraction, and subtract the coincidence background. This value should be dominated by the specular component for all the samples. The wafer's value is used for the absolute normalization. The uncertainty of $R_{\mathrm{spec}}$ is found by propagating the errors on the individual contributions (statistical errors on $\mu_{\mathrm{D}}$, background, light stability and containment corrections). Figure~\ref{fig:rspec_vs_aoi} shows the angular dependence of the specular reflectance in LXe for the three samples and wafer. 
The $R_{\mathrm{spec}}$ values for representative low, middle, and high angles are tabulated in Table~\ref{tab:rspec}. 
\begin{table}[htpb]
\centering
\begin{tabular}{| c| c| c| }
\hline
\textbf{Sample} & \textbf{Angle ($\degree$)} & \bm{$R_{\mathrm{spec}}$} (\%)\\  
\hline
 \multirow{3}{*}{Wafer} & 17 & 51 \\  
                        & 42 & 52.6$\pm$2.3 \\
                        & 74 & 85$\pm$17\\ 
 \hline
 \multirow{3}{*}{High polish} & 24 & 62.5$\pm$1.4 \\  
                        & 45 & 67.7$\pm$3.5 \\
                        & 66 & 80$\pm$8\\ 
 \hline
 \multirow{3}{*}{Average polish} & 20 & 27.6$\pm$0.8 \\  
                        & 40 & 40.7$\pm$2.1 \\
                        & 75 & 84$\pm$18\\ 
 \hline
 \multirow{3}{*}{Realistic} & 20 & 5.8$\pm$0.4 \\  
                        & 45 & 7.6$\pm$0.6 \\
                        & 75 & 23$\pm$5 \\ 
 \hline
\end{tabular}
\caption{$R_{\mathrm{spec}}$ in LXe of different samples at \textit{low}, \textit{middle} and \textit{high} angles.}
\label{tab:rspec}
\end{table}
As seen from the table, substrate polishing correlates strongly with $R_{\mathrm{spec}}$. Reflectance of all three samples increases with the angle of incidence, but the absolute value of $R_{\mathrm{spec}}$ for the realistic sample is still several times smaller than the target value of 80\% for all probed angles. It's clear from Figure~\ref{fig:rspec_vs_aoi} that a substantial fraction of light reflects non-specularly from the sample. Consequently, a total reflectance measurement is performed to understand whether the coating satisfies the requirement, as described in the next section.

\subsection{Reflectance in GN2}
\begin{figure*}[htpb]
    \centering
    \includegraphics[width=0.99\textwidth]{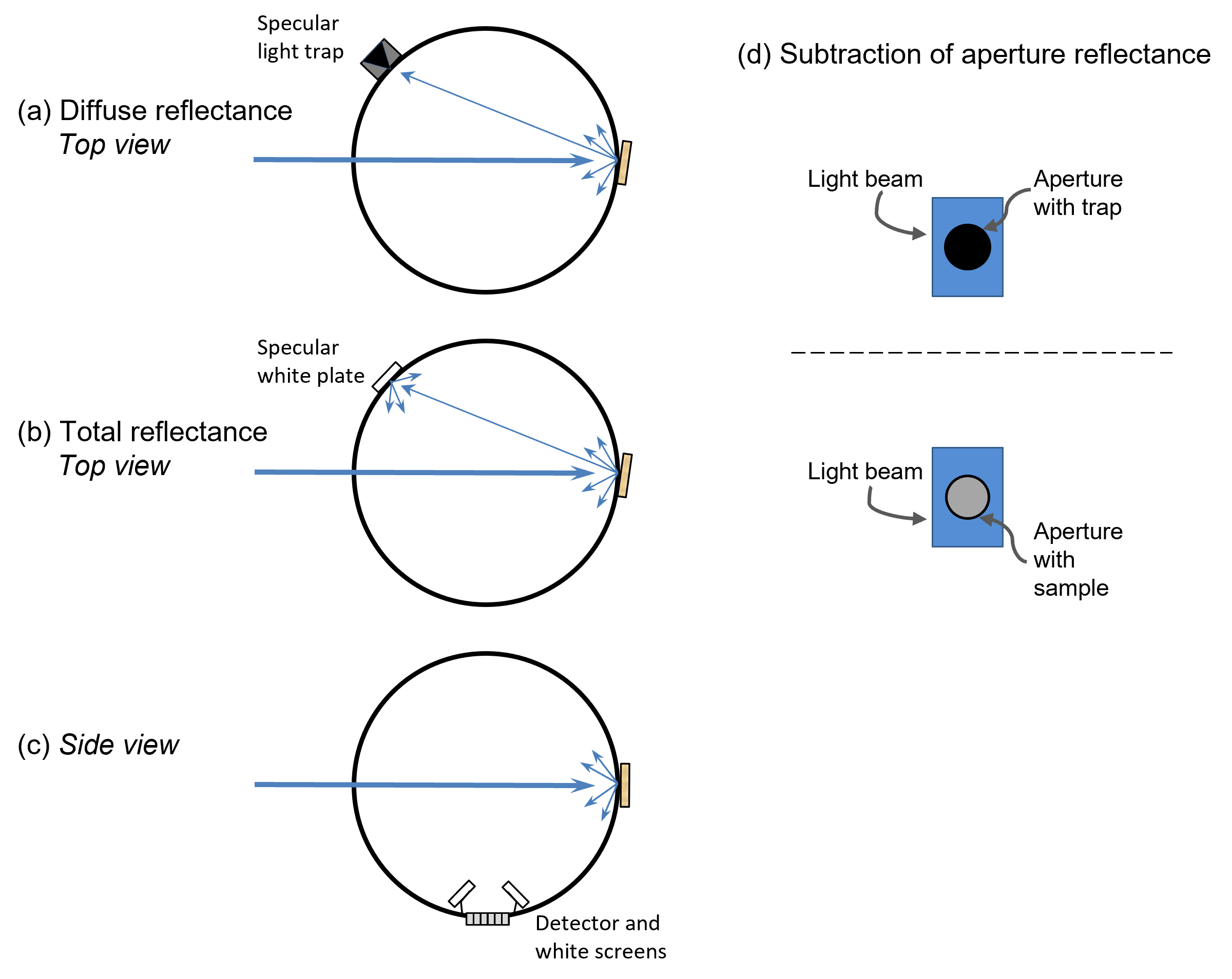}
    \caption{Integrating sphere measurement setup in GN2 (for clarity, the reference beam and reference beam port are not shown): (a) A light trap was used to remove the specular reflectance and measure the diffuse reflectance only, and was replaced by a white plate; (b) to measure the total reflectance including both specular and diffuse components. (c) The photomultiplier detector was positioned at the bottom of the sphere, and protected from directly viewing the sphere ports by Spectralon white screens. (d) Because the probing light beam area was larger than the sphere aperture for measuring the sample, a measurement without sample, with a second light trap, was subtracted from all measurements in order to remove the contribution of the light beam incident on the sphere wall around the aperture.} 
    \label{fig:sphere_setup}
\end{figure*}

The diffuse reflectance spectra of the samples were measured using a spectrophotometer (Perkin-Elmer Lambda-900) equipped with a 15-cm diameter integrating sphere (Labsphere), both purged with dry gaseous nitrogen (GN2) to reduce the oxygen absorption. The light source in the spectrophotometer is a deuterium lamp, and the sphere is equipped with a Hamamatsu R-955 photomultiplier tube detector, with an estimated quantum efficiency between 35\% and 40\% at 175 nm~\cite{rd955}. The sphere internal surface is covered with Spectralon, a pressed PTFE-like material developed by Labsphere~\cite{spectralon}. Figure~\ref{fig:sphere_setup} shows how the diffuse reflectance, $R_{\mathrm{diff}}$, was measured using a light trap (Avian Technologies, NH) to remove the specularly reflected signal. 
The total reflectance, $R_{\mathrm{tot}}$, was measured by replacing the light trap by a white plate (Spectralon). The $R_{\mathrm{spec}}$ could then be estimated as the difference between the total and diffuse components. To account for the incident light beam being larger than the sample aperture, a measurement of the aperture only -- with a light trap instead of a sample -- was subtracted from each measurement.

\begin{figure*}[htpb]
    \centering
    \includegraphics[width=0.99\textwidth]{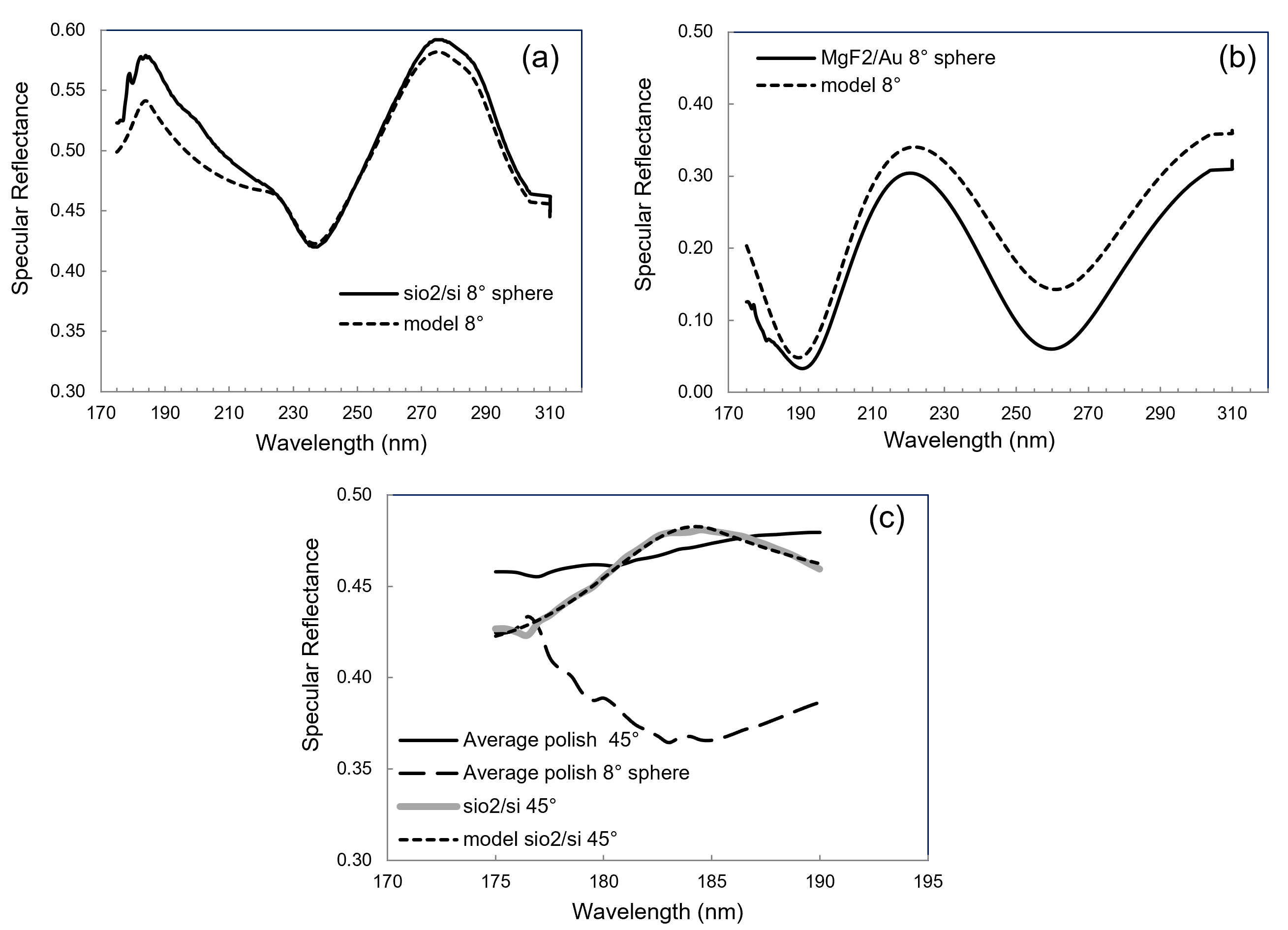}
    \caption{Comparison of specular reflectance spectra in GN2 deduced from integrating sphere measurements with prediction based on Fresnel equations and known VUV optical constants for the materials of the reference samples: (a) SiO$_2$ on a silicon wafer and (b) MgF$_2$ on gold mirror. (c) shows a comparison of specular reflectance spectra of the average polish sample as measured with the integrating sphere (at 8$\degree$ of incidence) and with a VN accessory (at 45$\degree$). Also shown is the reflectance spectra of the SiO$_2$/Si sample measured at 45$\degree$ and the corresponding prediction based on the Fresnel equations.} 
    \label{fig:specular_reflectance}
\end{figure*}

The wavelength of 175\,nm is at the lower limit of the Lambda-900 spectrophotometer and Labsphere integrating sphere's specifications. In order to increase the signal to noise ratio, a longer exposure time (10\,s, corresponding to 6\,nm/m) was used for the detector, and the spectrometer slit was kept relatively large, at 5\,nm.  For the reflectance measurements, a Spectralon sample was used as a reference. To ensure the validity of the results, it was critical that the Spectralon sample had no previous exposure to contamination or UV light, as the reflectance of this material is known to degrade significantly when contaminated~\cite{spectralon_degrad}. A new, out-of-the-box, Spectralon sample was used for this measurement. We estimate that the diffuse reflectance of the non-contaminated Spectralon at 175 nm was between 93\% and 95\%. It was done by comparing specular reflectance spectra measured at 8$\degree$ angle of incidence on a SiO$_2$/Si sample using the sphere with Spectralon as a reference material with specular spectra at 8$\degree$ and 45$\degree$ obtained using other reflectance accessories measuring directly the specular reflectance, and with thin-film optical models (see Figure~\ref{fig:specular_reflectance}(a) and (c)). 

\begin{figure*}[htpb]
    \centering
    \includegraphics[width=0.99\textwidth]{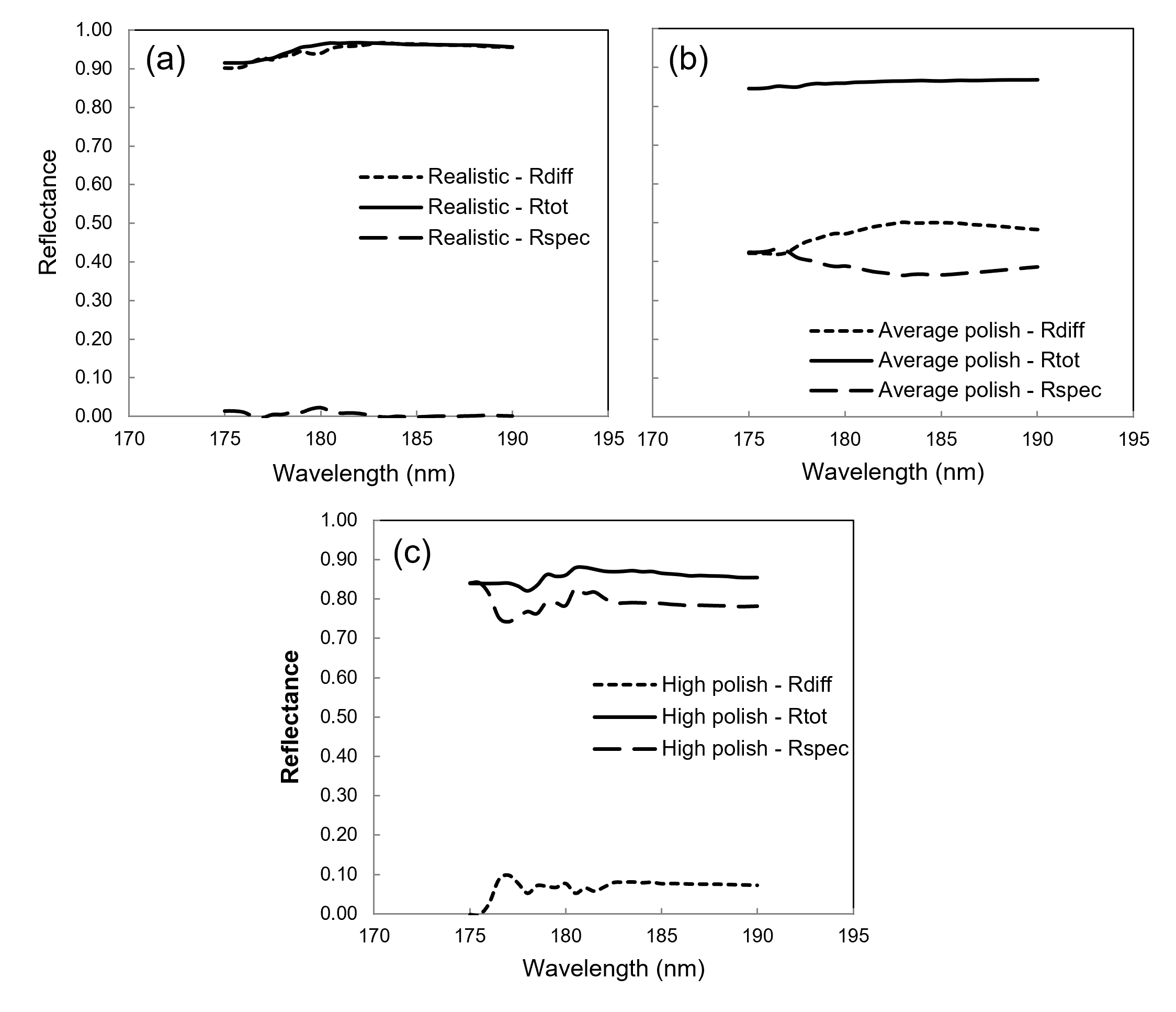}
    \caption{The diffuse, total, and specular reflectance spectra in GN2 for (a) realistic, (b) average polish, and (c) high polish samples. The specular reflectance is calculated as the difference between total and diffuse ones.} 
    \label{fig:diffuse_reflectance}
\end{figure*}

Figure~\ref{fig:diffuse_reflectance} shows the diffuse and total reflectance spectra of the realistic, average polish, and high polish samples, as measured with the integrating sphere.
As expected, the realistic sample shows a very low specular, but a high diffuse and total reflectances, as opposed to the high polish sample that has a high specular and a relatively low diffuse reflectances. The non-zero diffuse reflectance for the high polish sample may be explained by a few scratches of the surface. 

Table~\ref{tab:nrc_results} summarizes the results at the wavelength of interest, indicating that the total reflectance of all three samples satisfies the design requirements of the next-generation LXe experiments. 
\begin{table}[htpb]
\centering
\begin{tabular}{| c| c| c| c| }
\hline
\textbf{Sample} & \bm{$R_{\mathrm{tot}}$} (\%) & $R_{\mathrm{diff}}$ (\%)& $R_{\mathrm{spec}}$ at 8$\degree$ (\%)\\  
\hline
High polish   &  \textbf{84$\pm$8} & 0.0& 84.1 \\
\hline
Average polish  &  \textbf{85$\pm$8} & 42.2 & 42.4 \\
\hline
Realistic  &  \textbf{92$\pm$8} & 90.2 & 1.4 \\
 \hline
\end{tabular}
\caption{Reflectance values of the three samples as measured in GN2 (with the sphere) at 175\,nm.}
\label{tab:nrc_results}
\end{table}
The imperfect agreement of $R_{\mathrm{spec}}$ seen in GN2 (last column of Table~\ref{tab:nrc_results}) and in LXe (low angle entries in Table~\ref{tab:rspec}) can be attributed to the different incident medium and angle of incidence. 
Additionally, these results depend on the reflectance of the Spectralon reference sample that may be subject to degradation. To independently validate the results, additional sphere reflectance measurements were performed with SiO$_2$/Si and MgF$_2$-coated gold mirror~\cite{newport} samples, whose expected reflectance could be calculated using the Fresnel equations and known VUV optical constants~\cite{wvase}. As shown in Figures~\ref{fig:specular_reflectance}(a) and (b), the measured and predicted spectra are not identical, but close. The gold reflectance spectrum is also similar to that provided by the supplier~\cite{newport}. 
As another way to validate the sphere reflectance results, the specular reflectance spectra of several samples were measured in GN2 at 45$\degree$ with weakly polarized light (estimated 55$\degree$ polarization) using a VN specular reflectance accessory (with no reference sample required) and compared to results obtained with the sphere at 8$\degree$ and in LXe. Figure~\ref{fig:specular_reflectance}(c) shows the measured spectra for the SiO$_2$/Si sample and average polish sample. The spectra of the SiO$_2$/Si was modelled using the same optical model as in Fig.~\ref{fig:specular_reflectance}(a). The measured reflectance spectra of the average polish sample was compared to the 8$\degree$ specular reflectance spectra extracted from the sphere measurements. In addition, $R_{\mathrm{spec}}$ value at 175\,nm, $45.8$\%, was in reasonable agreement with the value from Table~\ref{tab:rspec}, $40.7$\%, measured in LXe at a similar angle of 40$\degree$.
Based on the above comparisons, we adopt 8\% abs. as a conservative uncertainty on the total reflectance measurements.

It is thus expected that the probable drop of reflectance of the new Spectralon reference sample in the VUV region would not affect significantly the results shown in Figure~\ref{fig:diffuse_reflectance}, which means that the total reflectance of the realistic sample should confidently be over 80\%.

\section{Concluding remarks}
\label{sec:conclusion}
The thin-film coatings studied in this work are found to posses satisfactory VUV reflectance for all three types of substrates. While the specular component of the realistic substrate is small, it is the total reflectance that matters for the next-generation experiments. This is predominantly because the size of the next-generation detectors will be much larger than the Rayleigh scattering length in LXe, effectively randomizing the photons' travel paths from the production point to the photosensors. The adhesion of the films to the substrates showed no visible signs of deteriorating after LXe measurements, during which the samples were cooled down to the LXe temperature (-100$\degree$C) for 10--20 hours. Overall, the performance of the thin-film coating under study is considered satisfactory for the next-generation LXe experiments, validating its design.

\acknowledgments
This work is supported by a Department of Energy Grant No. DE-SC0019261. We thank the PNNL's nEXO group for preparing the realistic substrates; the IOE and IHEP groups for preparing the samples. We gratefully acknowledge the support of Nvidia Corporation with the donation of three Titan Xp GPUs used for the optical simulations. I. Ostrovskiy thanks the Chinese Academy of Sciences (CAS) President's International Fellowship Initiative (PIFI) for the support.

\paragraph{Author Contributions.} W. Wang operated the LIXO2 system, conducted the LXe measurements, and performed low-level data analysis of the LIXO2 data. A. Best conducted optical simulations of the LIXO2 setup and participated in the LXe measurements. D. Bajpai participated in the LXe measurements, performed high-level analysis, and produced the LIXO2 results. D. Poitras conducted the GN2 measurements and produced their results, helped with writing and editing the manuscript. I. Ostrovskiy supervised the effort, wrote and edited the manuscript. All authors have read and agreed to the final version of the manuscript.

\bibliographystyle{JHEP} 
\bibliography{main}

\end{document}